\documentclass[pra,twocolumn,superscriptaddress,amsmath,amssymb,floatfi,showpacs]{revtex4}
\usepackage{graphicx}

\begin{document}
\title{\textbf{Giant vacuum forces via transmission lines}}
\author{Ephraim Shahmoon}
\affiliation{Department of Chemical Physics, Weizmann Institute of Science, Rehovot, 76100, Israel}
\author{Igor Mazets}
\affiliation{Vienna Center for Quantum Science and Technology, Atominstitut, TU Wien, 1020 Vienna, Austria}
\affiliation{Ioffe Physico-Technical Institute of the Russian Academy of Sciences, 194021 St. Petersburg, Russia}
\author{Gershon Kurizki}
\affiliation{Department of Chemical Physics, Weizmann Institute of Science, Rehovot, 76100, Israel}
\date{\today}

\begin{abstract}
Quantum electromagnetic fluctuations induce forces between neutral particles, known as the van der Waals (vdW) and Casimir interactions. These fundamental forces, mediated by virtual photons from the vacuum, play an important role in basic physics and chemistry, and in emerging technologies involving, e.g. micro-electromechanical systems or quantum information processing.
Here we show that these interactions can be enhanced by many orders of magnitude upon changing the character of the mediating vacuum-modes. By considering two polarizable particles in the vicinity of any standard electric transmission line, along which photons can propagate in one dimension (1d), we find a much stronger and longer-range interaction than in free-space. This enhancement may have profound implications on many-particle and bulk systems, and impact the quantum technologies mentioned above. The predicted giant vacuum force is estimated to be measurable in a coplanar waveguide line.
\end{abstract}
 \maketitle
The seminal works of London \cite{LON}, Casimir and Polder \cite{CP,CAS} identified quantum electromagnetic fluctuations to be the source of both short-range (vdW) and retarded, long-range (Casimir), interactions between polarizable objects, which may be viewed as an exchange of virtual photons from the vacuum. Subsequent studies of these interactions \cite{MIL,LAM,DEC,MOH,CAP,DAL,EMIG,REY,JOHN,MAZ,MILT} revealed their modifications, such as retardation and non-additivity \cite{REV}, brought about by the geometry of the polarizable objects. The ability to design these interactions is important for their possible use and exploration in emerging quantum technologies such as  micro-electromechanical systems \cite{ASK}, quantum information processing \cite{RYD,FIR} and circuit quantum electrodynamics, where the dynamical Casimir effect has been recently demonstrated \cite{WIL2,PAR}. Here we show that effectively one-dimensional (1d) transmission-line environments can induce strongly enhanced and longer-range van der Waals and Casimir interactions compared to free space. Such enhanced interactions may have profound implications on the quantum technologies mentioned above and give rise to a variety of new many-body phenomena involving polarizable particles in effectively 1d environments.

A key point in determining how these interactions depend on distance is the spatial propagation and scattering of the virtual photon modes that mediate them. Like any other waves, photons are scattered differently off objects with different geometries. For example, light is much more effectively scattered off a mirror than off a point-like atom. This explains the stronger and longer-range vacuum interaction between mirrors \cite{CAS}, compared to that between atoms \cite{CP}. This idea also underlies the dependence of Casimir forces on the geometrical shape of the interacting objects.
Here, we take a somewhat different approach towards the geometry dependence of van der Waals (vdW) and Casimir-related phenomena. Instead of considering the interaction energy of extended objects with different geometries, we revisit the original Casimir-Polder \cite{CP} configuration of a pair of point-like dipoles while changing the geometry of their surrounding environment such that it confines the propagation of virtual photons to a certain direction. More specifically, we consider the energy of the interaction between two dipoles, mediated by vacuum photon modes along a 1d transmission line (TL). The resulting attractive interaction is found to be much stronger and longer-range than its free-space counterpart. Surprisingly, this interaction scales with the inter-dipolar distance $r$ as $const.+(r/\lambda) \ln(r/\lambda)$ or as $1/r^3$ for shorter or longer $r$ than the typical dipolar wavelength $\lambda$, respectively, as opposed to the corresponding $1/r^6$ or $1/r^7$ scalings in free space \cite{CP,MIL}. This enhancement implies a drastic modification of Casimir-related effects for many-body and bulk polarizable systems in such an effectively 1d geometry.

This article is organized as follows: The \emph{Analysis Principles} section presents the analytical approaches used to obtain the van der Waals and Casimir interactions in a TL environment, Eqs. (\ref{E}) and (\ref{F}). The \emph{Predictions} section reveals the giant enhancement of these interactions by comparing them to the case of free-space [Figs. 3(c) and 3(d)], considering possible experimental realizations and imperfections. The consequences of this giant interaction to generalized Casimir effects in 1d is addressed to in the \emph{Prospects} section, followed by our \emph{Conclusions}.

\section*{ANALYSIS PRINCIPLES}

\subsection*{1d photons in transmission lines}
The ability to change the geometry of photon vacuum modes is widely used in quantum optics, e.g. for the enhancement of spontaneous emission \cite{KOF,SPON} and resonant dipole-dipole interactions \cite{RDDI}.
Here, effectively 1d propagation of the virtual photons is attained in waveguide structures that support \emph{transverse-electromagnetic} (TEM) modes, namely, modes whose propagation axis, the electric and the magnetic fields, are perpendicular to each other. These are typically the fundamental transverse modes of electric transmission lines (TLs), i.e. waveguides comprised of two conductors as shown in Fig. 1, which constitute the standard workhorse of electronic signal transmission. They possess a dispersion relation $\omega_k=|k| c$ and an electric field-mode function \cite{POZ}
\begin{equation}
\mathbf{u}_{kj}(\mathbf{r})=\frac{1}{\sqrt{A(x,y)L}}e^{i k z}\mathbf{e}_j,
\label{u}
\end{equation}
where $z$ is the propagation axis, $L$ its corresponding quantization length, $A(x,y)$ the effective area in the transverse $xy$ plane, $\mathbf{e}_{j=x,y}$ the polarization unit vector, $k$ and $\omega_k$ the wavenumber in the $z$ direction and the angular frequency, respectively, and  $c=1/\sqrt{\mu\epsilon}$ the phase velocity, $\mu$ and $\epsilon$ being the effective permeability and permittivity of the TEM mode, set by the geometry and materials of the TL.

Unlike modes of other waveguides, such as optical fibers or hollow metallic waveguides, here the effective area $A(x,y)$ is independent of frequency. Hence, considering its dispersion relation, the unique feature of the TEM mode is that it forms an effective plane wave in 1d.
This special property enables the TEM mode to guide or confine virtual photons over much longer distances than all other higher-order transverse modes supported by the TL, such as the transverse electric (TE) and transverse magnetic (TM) modes in a coaxial TL \cite{POZ}. As recently shown by us for metallic waveguides \cite{MWG}, such modes do not contribute to the long-range interaction (Materials and Methods).

Having identified the key role of the TEM mode in mediating vacuum interactions in a TL, we proceed to the analytically calculate the vdW/Casimir interaction between a pair of dipoles mediated by the TEM mode using two different methods (two next  subsections) and present the results (last subsection).

\begin{figure}
\begin{center}
\includegraphics[scale=0.3]{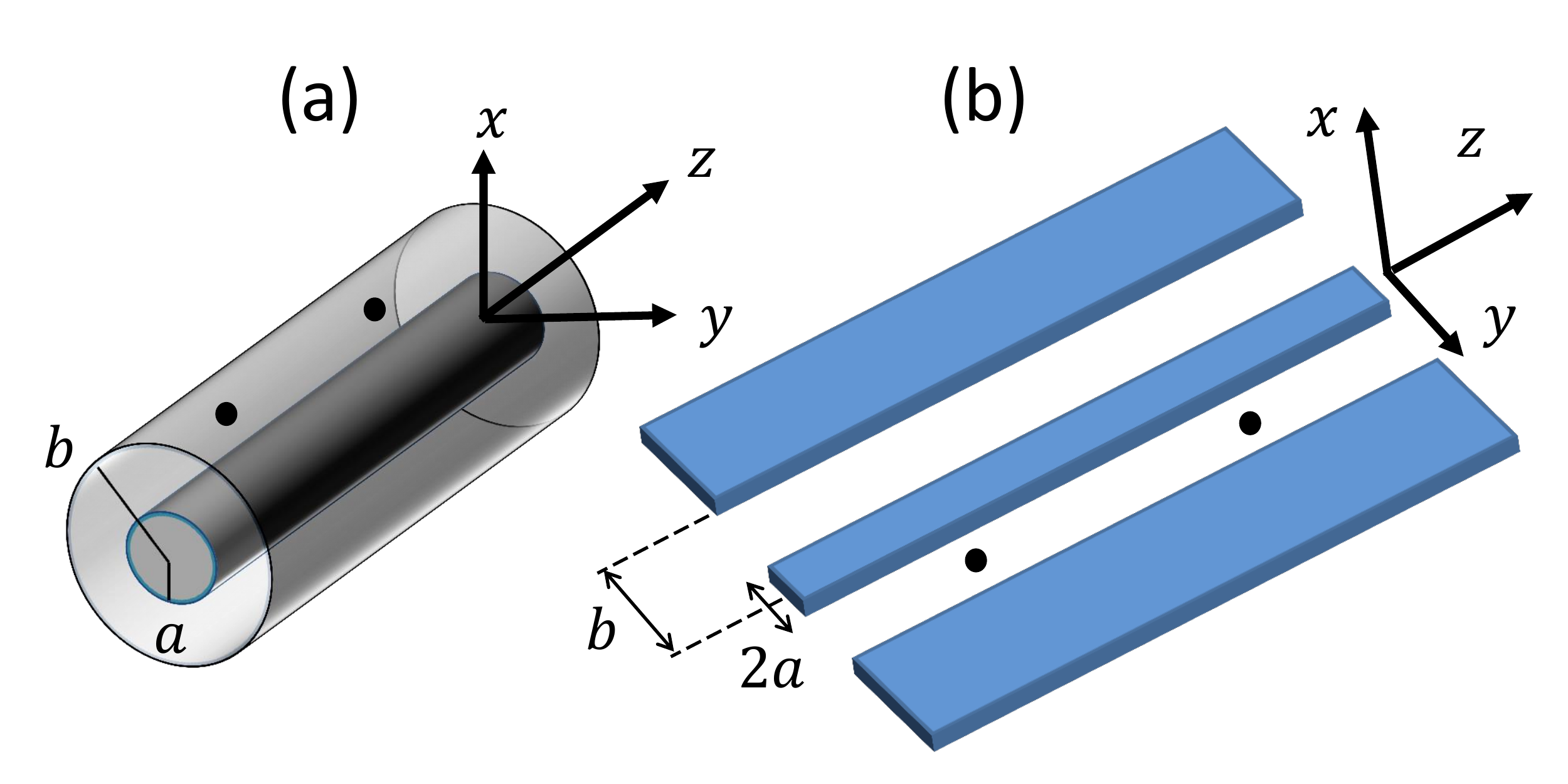}
\caption{\small{Geometries of transmission-line mediated vdW and Casimir interactions. (a) Coaxial line: two concentric metallic cylinders, the inner one with radius $a$ and the outer (hollow) one with radius $b$. Two dipoles represented by black dots are placed inbetween the cylinders, along the wave propagation direction $z$. They interact via modes of the coaxial line that are in the vacuum state, giving rise to a vdW-like interaction energy. (b) Coplanar waveguide: similar to (a). Here the central conductor of width $2a$ is separated from a pair of ground plane conductors that are $2b$ apart.
}}
\end{center}
\end{figure}

\subsection*{QED perturbation theory}
We first adopt the quantum electrodynamics (QED) perturbative approach of Ref. \cite{MQED}, to two identical atomic or molecular dipoles with a ladder of excited levels $\{|n\rangle\}$, both in their ground state $|g\rangle$, that are coupled to the vacuum of photon modes given by equation (\ref{u}), via the interaction Hamiltonian
\begin{equation}
H_I=-\hbar\sum_{\nu=1}^2\sum_{n_{\nu}}\sum_{k,j}\left(|n_{\nu}\rangle \langle g|+\mathrm{h.c.}\right)\left(i g_{kj\nu n} \hat{a}_{kj} +\mathrm{h.c.}\right).
\label{H}
\end{equation}
Here $g_{kj\nu n}=\sqrt{\frac{\omega_k}{2\epsilon\hbar}}\mathbf{d}_{n_{\nu}}\cdot\mathbf{u}_{kj}(\mathbf{r}_{\nu})$ is the dipolar coupling between the dipole $\nu=1,2$ and the $kj$ field mode, for the $|g\rangle$ to $|n_{\nu}\rangle$ transition with dipole matrix element $\mathbf{d}_{n_{\nu}}$, $\hat{a}_{kj}$ being the lowering operator for the $kj$ mode. The vdW energy is then obtained by perturbation theory as the fourth-order correction to the energy of the ground state $|G\rangle=|g_1,g_2,0\rangle$, where $|0\rangle$ is the vacuum of the photon modes \cite{MQED},
\begin{equation}
U=-\sum_{I_1,I_2,I_3}\frac{\langle G|H_I|I_3\rangle \langle I_3|H_I|I_2\rangle \langle I_2|H_I|I_1\rangle \langle I_1|H_I|G\rangle}{(E_{I_1}-E_G)(E_{I_2}-E_G)(E_{I_3}-E_G)}.
\label{E4}
\end{equation}
Here $|I_j\rangle$ are intermediate (virtual) states (of the free Hamiltonian) and $E_m$ is the energy of the state $|m\rangle$.
The sum in equation (\ref{E4}) includes 12 different terms, each corresponding to a different set of virtual processes and represented by a diagram, e.g. that of Fig. 2(a) \cite{MQED}. Each of the 12 terms contains a summation over all the dipolar states $|n_1\rangle$ and $|n_2\rangle$, and the photonic polarizations $j,j'$, and integrations over the wavevectors $k$ and $k'$.
The energy is then obtained by summing all 12 terms and performing the integrations (Materials and Methods).

 \begin{figure}
\begin{center}
\includegraphics[scale=0.30]{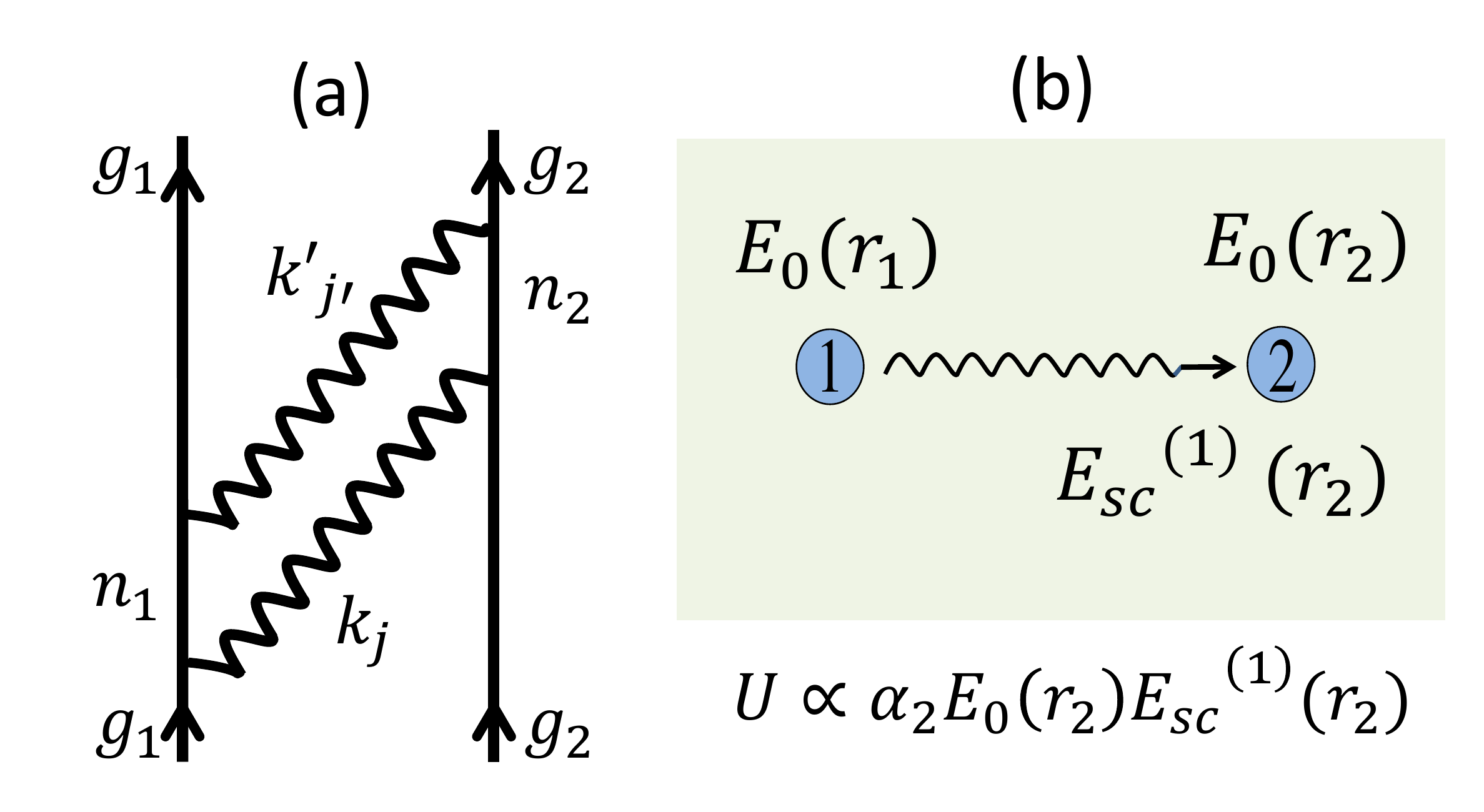}
\caption{\small{Calculation methods of the interaction energy.
(a) QED perturbation theory: one of 12 possible processes (diagrams) that contribute to the energy correction of the state $|G\rangle=|g_1,g_2,0\rangle$, equation (\ref{E4}). Here the intermediate states are $|I_1\rangle=|n_1,g_2,1_{kj}\rangle ,\, |I_2\rangle=|g_1,g_2,1_{kj}1_{k'j'}\rangle$ and $|I_3\rangle=|g_1,n_2,1_{k'j'}\rangle$, where $|1_{kj}\rangle=\hat{a}^{\dag}_{kj}|0\rangle$.
(b) Scattering of vacuum fluctuations: the vacuum field $\hat{\mathbf{E}}_0(\mathbf{r})$ exists in all space and interacts with both dipoles at their positions $\mathbf{r}_1$ and $\mathbf{r}_2$, hence it is also scattered by the dipoles. The scattered field from dipole 1, $\hat{\mathbf{E }}^{(1)}_{sc}(\mathbf{r})$, arrives at dipole 2, resulting in an interaction interpreted as the vdW/Casimir interaction $U$.
}} \label{QED}
\end{center}
\end{figure}

\subsection*{Scattering of vacuum fluctuations}
A more transparent approach is based on the solution of the 1d electromagnetic wave equation "driven" by the vacuum field  \cite{MIL}. Accordingly, we consider two point dipoles with polarizabilities $\alpha_{1,2}(\omega)$, subject to a fluctuating (vacuum) field $\hat{\mathbf{E}}_0$ [Fig. 2(b)]. The electromagnetic energy of, say, dipole 2, is given by a sum over all $k$ of $-(1/2)\alpha_2(\omega_k)[\hat{\mathbf{E}}_k(\mathbf{r}_2)\hat{\mathbf{E}}_k^{\dag}(\mathbf{r}_2)+\hat{\mathbf{E}}_k^{\dag}(\mathbf{r}_2)\hat{\mathbf{E}}_k(\mathbf{r}_2)]$, where $\hat{\mathbf{E}}_k(\mathbf{r}_2)$ is the $k$ mode component of the electromagnetic field at the location of this dipole. The field at $\mathbf{r}_2$ includes two components: $\hat{\mathbf{E}}_{0,k}(\mathbf{r}_2)$, the external fluctuating (vacuum) field, and $\hat{\mathbf{E}}^{(1)}_{sc,k}(\mathbf{r}_2)$, the scattered field, which to lowest order is that scattered from dipole 1, where it is driven by the vacuum fluctuations $\hat{\mathbf{E}}_{0,k}(\mathbf{r}_1)$ at $\mathbf{r}_1$. Since the TEM field exhibits diffraction-less propagation in 1d, this scattered field is found by essentially solving
\begin{equation}
\left(\partial_z^2+k^2\right)\hat{\mathbf{E}}^{(1)}_{sc,k}(z)=-\mu \omega_k^2 \alpha^{(1d)}_1(\omega_k)\hat{\mathbf{E}}_{0,k}(z_1)\delta(z-z_1),
\label{HELM}
\end{equation}
where the quantum (vacuum) field in 1d is $\hat{\mathbf{E}}_{0,k}(z)=i\sqrt{\frac{\hbar\omega_k}{2\epsilon}}\frac{1}{\sqrt{L}}e^{ikz}\mathbf{e}_j \hat{a}_{kj}$ and $\alpha_{\nu}^{(1d)}=\alpha_{\nu}/A(x_{\nu},y_{\nu})$. The vdW/Casimir energy is then obtained as the interaction energy between the dipoles, which to lowest order becomes
\begin{eqnarray}
&&U=-2\sum_k\alpha^{(1d)}_2(\omega_k)
\nonumber\\ &&\times
\langle0|\left[\hat{\mathbf{E}}^{(1)}_{sc,k}(z_2)\hat{\mathbf{E}}_{0,k}^{\dag}(z_2)+\hat{\mathbf{E}}_{0,k}(z_2)\hat{\mathbf{E}}_{sc,k}^{(1)\dag}(z_2)\right]|0\rangle.
\label{U}
\end{eqnarray}
Here we average over the quantized fields in the vacuum state $|0\rangle$, and multiply by 4 to account for both field polarizations $j$ and for the energy at the locations of both dipoles.

\subsection*{Analytical results}
Both calculation methods yield the same result for the TEM-mediated interaction energy between dipoles with excited levels $\{|n\rangle\}$ (Materials and Methods). In the main text, we only present the resulting energy for the case where one excited level $|e\rangle$ with energy $E_e$ and corresponding wavelength $\lambda_e=2\pi\hbar c/E_e$, has a dominant dipole transition, such that all other excited levels $|n\neq e\rangle$ can be neglected. This yields
\begin{eqnarray}
U(z)&=&-\frac{\pi|\mathbf{d}_e^{\bot}|^4}{2\epsilon^2 E_e} \frac{1}{A_1 A_2\lambda_e^2}F(z),
\nonumber \\
F(z)&=&\left(4\pi\frac{z}{\lambda_e}+i\right)e^{-i4\pi\frac{z}{\lambda_e}}\mathrm{Ei}(i4\pi\frac{z}{\lambda_e})
\nonumber\\
&&+\pi\left(1+i4\pi\frac{z}{\lambda_e}\right)e^{i4\pi\frac{z}{\lambda_e}}+\mathrm{c.c.},
\label{E}
\end{eqnarray}
where $z=|z_2-z_1|$, is the inter-dipolar distance in the TL propagation direction, $A_{\nu}=A(x_{\nu},y_{\nu})$, $\mathbf{d}_e^{\bot}$ is the projection of the dipole matrix element $\mathbf{d}_e$ on the transverse $xy$ plane, $\mathrm{c.c.}$ stands for complex-conjugate and $\mathrm{Ei}(x)=-\mathrm{P}\int_{-x}^{\infty}dt e^{-t}/t$ is the exponential integral function. This interaction is attractive and its dependence on distance $z$ is described by the monotonously decreasing function $F(z)$, plotted in Figs. 3(a) and 3(b). From $F(z)$ we obtain the energy dependence  for small and large $z/\lambda_e$, respectively,
\begin{eqnarray}
&&F(z\ll \lambda_e)\approx \pi+16 \pi\frac{z}{\lambda_e} \ln \frac{z}{\lambda_e}+ 8\pi[2\gamma -1 + 2\ln(4\pi) ]\frac{z}{\lambda_e},
\nonumber\\
&&F(z\gg \lambda_e)\approx \frac{1}{8\pi^3}\frac{\lambda_e^3}{z^3},
\label{F}
\end{eqnarray}
where $\gamma\cong 0.577$ is Euler's constant. These expressions for $F(z)$ in the vdW, $z\ll \lambda_e$, and Casimir (retarded), $z\gg \lambda_e$, regimes reveal a universal scaling with distance $z$; namely, it does not depend on $\lambda_e$ or any other length-scale, but rather on $z \ln z$ and $1/z^3$, respectively. This scaling is thus independent of the type of dipole involved in the interaction and is expected to hold for multilevel dipoles as verified by Eqs. (11) and (\ref{F12b}) in the Materials and Methods section: The level structure merely affects the prefactor, i.e. the coefficient of the interaction.

\section*{PREDICTIONS}
The expressions given above for the interaction energy in the vdW and Casimir regimes, suggest the possibility of a much stronger interaction than its free-space counterpart. In the nonretarded vdW regime, it decreases very slowly with $z$ compared to the familiar $1/z^6$ scaling, whereas in the retarded Casimir regime, it falls off with a power-law which is four powers weaker than its $1/z^7$ Casimir-Polder counterpart.

\subsection*{Quantitative comparison to free-space vacuum forces}

Let us make this comparison more quantitative by considering a general TL with separation $a$ between its two guiding conductors (Fig. 1). The effective TEM mode area $A(x,y)$ then scales as $a^2$ (Materials and Methods). Moreover, assuming the polarizability is isotropic on average, we can estimate $|\mathbf{d}_e^{\bot}|^2=(2/3)|\mathbf{d}_e|^2$. Then, for $\sqrt{A_{1,2}}=a$ and $a/\lambda_e\sim 10^{-4}$, typical of the circuit QED realizations considered below, where $a\sim$ few $\mu$m and $\lambda_e\sim$ few cm, the ratio between the TEM-mediated energy and its free-space counterpart is plotted in Figs. 3(c) and 3(d) for short and long distances, respectively \cite{MQED}. At distances longer than $z=10^{-3}\lambda_e$, the TEM-mediated interaction is enhanced by orders of magnitude compared to free space, and the enhancement factor increases drastically with $z$.

 At very short distances, however, the free-space vdW interaction $1/z^6$ is stronger. This occurs at $z<a$, where the dipoles are close enough so that they do not "sense" the TL structure. The transition to free-space behavior in this regime can be described by including higher-order transverse modes. These modes become significant at such short distances, where their contributions sum up to give the free-space result (Materials and Methods).

\subsection*{Possible experimental realizations}
It is important to consider the possibilities of measuring the predicted effects. Here we focus on the coplanar waveguide (CPW) TL [Fig 1(b)] that is extensively used in the emerging field of circuit-QED \cite{SCH2,WAL1,WAL2}. We consider the dominant dipole transition between the ground and first excited states of a pair of superconducting transmons, where other dipolar transitions from the ground state are indeed negligible \cite{SCH2} such that the two-level approximation is valid. The transmons, at distance $z$, are both capacitively coupled to the CPW. Then, the TEM-mediated interaction should induce a $z$-dependent energy shift on the dipole-transition levels, $U(z)$ as per equation (\ref{E}). we estimate this energy shift using the parameters of a recent experiment \cite{WAL1}: the dipole frequency is $E_e/(2\pi\hbar)\sim5$GHz and the dipole coupling to a closed CPW cavity of length $\lambda_e$, can reach $g\sim\pi\times720$MHz. From the relations $g=\sqrt{\frac{(E_e/\hbar)}{2\epsilon \hbar}}\frac{|\mathbf{d}_e|}{\sqrt{AL}}$ and $L=\lambda_e$ \cite{WAL1,WAL2}, we can extract the factor $|\mathbf{d}_e^{\bot}|/\sqrt{A}$ for each dipole, obtaining $[U(z)/F(z)]/h\sim 0.84$MHz, where $h$ is Planck's constant. E.g., for distances $z=0.001\lambda_e$ or $0.01\lambda_e$ (both much larger than $a\sim$few $\mu$m), the energy shift becomes $U(z)/h\approx 1.8$ or $2.47$MHz, respectively, about twice the dephasing rate of the dipole, $1$MHz \cite{WAL1}, that limits the resolution of the shift. This resolution can be considerably improved, as in Ref. \cite{WAL2}, where a dephasing time of about $20\mu$s is reported. Then, upon taking $E_e/h\sim2$GHz with the parameters of Ref. \cite{WAL1}, one can obtain $U(z)/h\approx 28$MHz for $z=0.01\lambda=1.5$mm, which is much larger than the dephasing rate.

Probing the interaction in the retarded regime, where $z>\lambda_e$, is currently more challenging: for the latter case, with $z=2\lambda_e=30$cm, we obtain $U(z)/h\approx 6.62$KHz, which is currently too small to be observed. Nevertheless, these results imply the remarkable possibility to directly observe the vdW and Casimir interaction of a \emph{single} pair of point-like dipoles (here the size of the dipoles is $\sim1-100\mu$m $\ll\lambda_e$) over a wide range of distances where the interaction scales non-trivially according to equation (\ref{E}).

We note that the ratio of the TL-length scale $a$ to $\lambda_e$ does not affect the scaling of the ratio $U/U_{fs}$ with $z$, but only its prefactor [see equation (\ref{E})]. Hence, the same scaling of $U/U_{fs}$ is expected for atoms, although their dipolar $\lambda_e$ can become much smaller than that of the superconducting transmon. For instance, considering the D1 line of Rb87 atoms, where $\lambda_e\approx 780$nm, and keeping the same TL with $a\sim 1$ $\mu$m, we obtain $a/\lambda_e>1$ (instead of $a/\lambda_e\sim 10^{-4}$ in Figs. 3c and 3d). The plots of Figs. 3c and 3d then retain the same slope but are shifted down. Namely, an enhancement, $U/U_{fs}>1$, still exists, but at larger distances $z$, where the enhanced interaction may be too weak to be observed for a single pair of atoms. Nevertheless, this enhancement can be very important for \emph{many-atom} systems where it would increase the non-additivity of their dispersion interactions (see Prospects below).

\begin{figure}
\begin{center}
\includegraphics[scale=0.2]{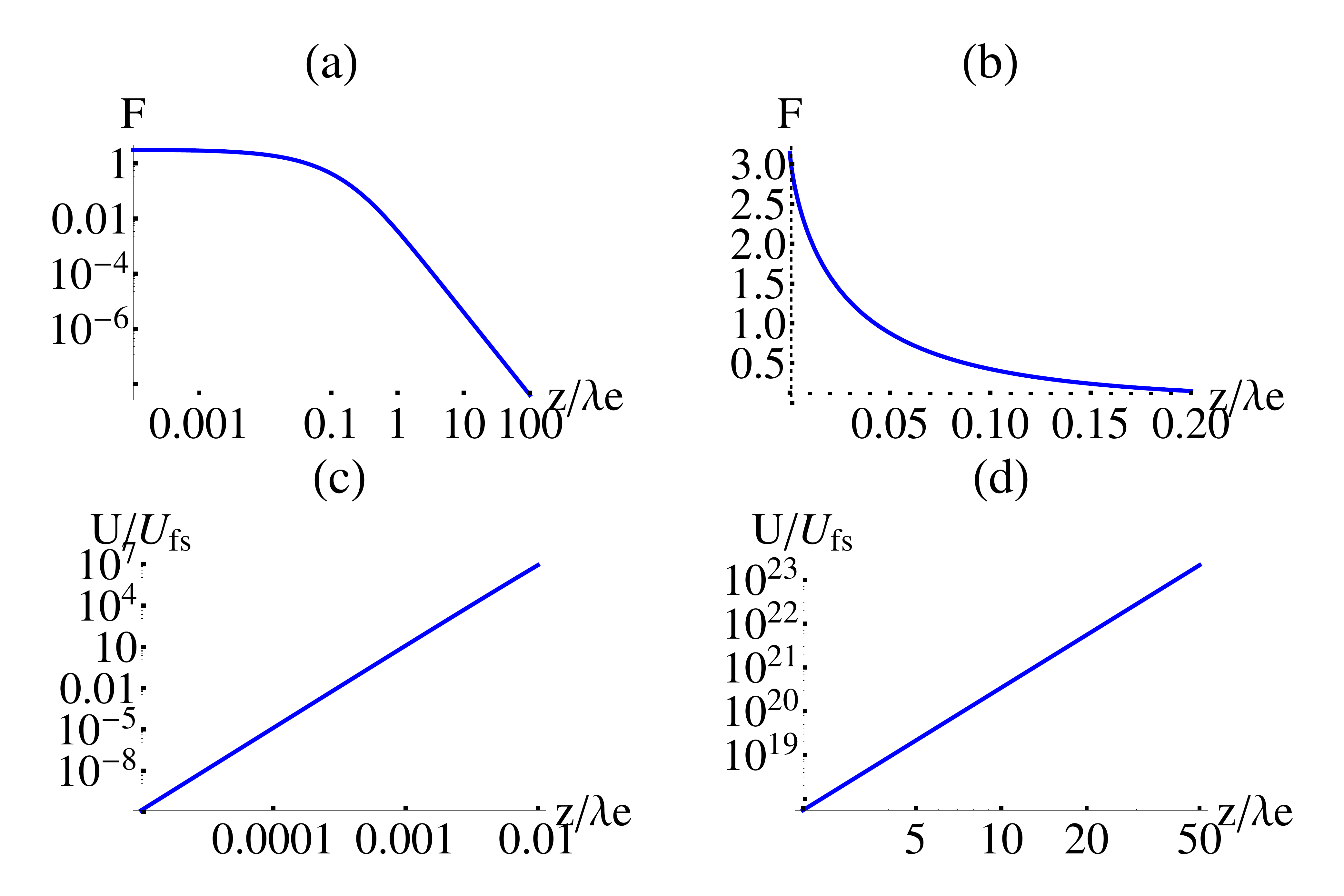}
\caption{\small{The TEM-mode mediated interaction potential as a function of inter-dipolar distance $z$.
(a) Log-log plot of $F(z)$, equation (\ref{E}). For long distances, $z\gg\lambda_e$, the linear dependence implies a power-law behavior as in equation (\ref{F}).
(b) $F(z)$ at short distances.
(c) Log-log plot of the ratio between the TEM-mediated energy $U(z)$, equation (\ref{E}), and the free-space vdW energy at short distances $z\ll\lambda_e$, $U_{fs}(z)=-\frac{|\mathbf{d}_e|^4}{48\pi^2\epsilon^2 E_e}\frac{1}{z^6}$ \cite{MQED} , with $|\mathbf{d}_e^{\bot}|^2=(2/3)|\mathbf{d}_e|^2$ and $\sqrt{A_{1,2}}=a$ (see text). Here $a=10^{-4}\lambda_e$, consistent with typical cases considered below, where $a\sim$few $\mu$m and $E_e/(2\pi\hbar)\sim$few GHz. Beyond $z\sim10^{-3}\lambda_e$, the huge enhancement of the interaction w.r.t its free-space counterpart is apparent.
(d) Same as (c), but for long distances $z\gg\lambda_e$ where the free-space energy takes the Casimir-Polder form, $U_{fs}(z)=-\frac{23}{64 \pi^3} \frac{\hbar c}{\epsilon^2}\frac{4}{9}\frac{|\mathbf{d}_e|^4}{E_e^2}\frac{1}{z^7}$ \cite{MQED}.
}} \label{f}
\end{center}
\end{figure}
\subsection*{Imperfections}
Let us consider the consequences of possible imperfections in the dipoles or the conductors of the TL.

\emph{Dipoles}.---
The description of sharp energy levels fits well the case of atoms but not  artificial dipolar systems such as transmons or quantum dots, for which \emph{non-radiative dephasing} may result in level widths. This dephasing may give a lower bound for measurable vdW/Casimir interaction-induced shifts, as discussed  above for the circuit QED realization. Another imperfection for artificial dipoles is their \emph{inhomogeneity}, e.g. that the dipole matrix element $\mathbf{d}_e$ and the energy $E_e$ of the dipolar transition in two different transmons are not identical. This does not change the scaling of the vdW interaction $U(z)$; however, it requires to  first measure the transmons' parameters if one wishes to obtain an exact result for $U(z)$ [see Eqs. (11)-(\ref{F12b}) and Supporting Information].

\emph{TLs}.---
The existence of TEM modes as in equation (\ref{u}) is exactly correct for TLs made from perfect conductors. However, it is also an excellent approximation to the propagation of fields in \emph{finite-conductivity}, low-loss, TLs that are currently used  \cite{POZ}, particulary in the microwave $\sim$GHz domain that is relevant in our case. This supports the validity of our results for TEM-mediated interaction in Eqs. (\ref{E}) and (\ref{F}) in realistic circuit QED systems. The conductors of the TL can also induce a \emph{modified Lamb shift} on dipole levels due to the photon modes near a metal surface, the so-called Casimir-Polder atom-surface interaction \cite{CP,HEN1}. This \emph{single-dipole} (rather than \emph{interaction-related}) energy shift is of no interest here, yet it may change $E_e$. Whereas for artificial dipoles $E_e$ is not precisely known anyway due to inhomogeneity, so that this effect is unimportant, the evaluation of this shift for atoms is described in the Supporting Information.
\emph{Surface roughness} along the TL conductors may slightly change the transverse profile of the TEM mode $1/\sqrt{A(x,y)}$; however, the 1d-like TEM mode, and hence the 1d vdW/Casimir interaction discussed above, still exists. \emph{Impurities and sharp edges} in the conductors can be treated as scatterers characterized by a reflectivity or polarizability, whose interaction with the dipoles contributes to the dipoles' energy shift.

\section*{PROSPECTS: CASIMIR PHYSICS IN 1D}
The vacuum force between two point dipoles underlies other vdW- or Casimir-related phenomena. Hence, the giant enhancement and non-trivial scaling of these forces with distance found here may open the door to unexplored Casimir-related effects in 1d, upon extending the present results to either multiple point-like dipoles or to extended (bulk) objects, as detailed in what follows.
\subsection*{Many-particle systems}
Since the open geometry of the CPW allows for the coupling of the TEM mode to clouds of trapped atoms above its surface \cite{MAY},
the predicted long-range interaction may be explored in a many-body setting.
This may entail a modification of the \emph{non-additivity} of the vdW and Casimir interactions, which currently attracts considerable interest \cite{REV}. In free-space, the vdW energy of a gas is approximately additive, namely, it can be obtained by pairwise summation of the vdW energies of all pairs, as long as $\alpha/r^3$ is small, $r$ being a typical inter-dipolar separation \cite{MIL}. This scaling is however a direct consequence of dipole-dipole interactions in free space, and is expected to change when atomic dipoles interact via the TEM mode of the CPW, for the same reasons that the $1/r^6$ scaling was shown here to change. Such atomic clouds are anticipated to exhibit non-additive effects at smaller densities than usual, which in turn may influence their dielectric properties, that may deviate from those obtained by the Clausius-Mossotti formula \cite{BW}. Furthermore, it would be interesting to consider how the enhanced interaction we predict is modified at finite temperatures or out of equilibrium \cite{THI,SHER,SCH}.

From a more applied point of view, the van der Waals energy shift between (out-of-equilibrium) Rydberg atoms, which underlies the blockade mechanism  used to design quantum gates \cite{RYD}, may be enhanced when the atoms are coupled to a TL. For Rydberg atoms ($\lambda_e\sim 1$cm) in free-space, blockade distances of $\sim 10\mu$m have already been observed \cite{FIR}, and the long-range scaling expected for their TL-mediated interaction may lead to the extension of the blockade distance even further.

\subsection*{Extended (bulk) objects}
Another direction to explore is the modification of the Casimir interaction involving bulk objects in a TL environment, and its possible relevance for actuation of micro-electromechanical systems \cite{ASK}. For example, one could consider the interaction energy between a dipole and a mirror \cite{CP,MIL}, where the role of the mirror may be played by a short-circuit at one of the ends of the TL. A mirror, whose reflection is not necessarily perfect, could be realized by an impedance characterized by capacitance and inductance.

In fact, in 1d there is no difference between a point dipole and an imperfect mirror. As shown above, for distances $z>a$ between the interacting objects, the TEM mode is dominant and a TL environment is effectively 1d. Hence, the theory of the Casimir interaction between mirrors in 1d \cite{REY2} finds a realistic context in TL environments. It is therefore interesting to compare its results to ours. The Casimir force between two mirrors in 1d, with frequency-dependent reflection coefficients $r_{1,2}(k)$ $(\omega=kc)$ is \cite{REY2}
\begin{equation}
f(z)=\int_{0}^{\infty}\frac{d k}{2\pi}\hbar c k\frac{ -r_1(k)r_2(k)e^{i2 k z}}{1-r_1(k)r_2(k)e^{i2 k z}}+\mathrm{c.c.}
\label{REY}
\end{equation}
For the case of dipoles, that we assume to be weak scatterers, we take $r_1r_2\ll 1$. The integral for the energy $U=-\int dzf(z)$ over imaginary frequencies $k=iu$ (Wick rotation, see Materials and Methods) then becomes,
\begin{equation}
U(z)\approx \frac{\hbar c}{2\pi}\int_0^{\infty}dur_1(iu)r_2(iu)e^{-2uz}.
\label{U1d}
\end{equation}
In the Casimir-Polder limit of large separations $z$, upon taking $\omega\rightarrow0$ \cite{MQED,MIL} and hence replacing $r_{1,2}(iu)$ with $r_{1,2}(0)$, we obtain $U\propto-r_{1}(0)r_{1}(0)/z$. The $1/z$ dependence is shown in \cite{REY2} to be universal and hold for any magnitude of the reflectivity $r_{1,2}$. This is in contrast with our $U\propto-1/z^3$ scaling [Eq. (\ref{F})]. The latter can be recovered from (\ref{U1d}) upon taking the limit $\omega\rightarrow0$ more carefully: From the Helmholtz equation (\ref{HELM}) we obtain that the reflection coefficient of the dipole, defined by $\hat{\textbf{E}}_{sc,k}^{(1)}=r_1(\omega_k)\hat{\textbf{E}}_{0k}$, becomes $r_{1,2}(\omega)\propto \omega \alpha_1(\omega)$. Our results are then retrieved from the integral (\ref{U1d}) [see Eq. (\ref{UABW})]. In particular, taking the Casimir-Polder limit $r_{1,2}(\omega\rightarrow0)\propto \omega \alpha_{1,2}(0)$ in (\ref{U1d}), we obtain $U\propto-1/z^3$ as in Eq. (\ref{F}).
This shows that the scaling of the energy $U$ with distance $z$ strongly depends on the frequency dependence of the reflectivities near $\omega\rightarrow 0$. E.g. for $r_{1,2}(\omega\rightarrow 0)\propto \omega^p$, we obtain in the retarded Casimir-Polder limit $U\propto-1/z^{2p+1}$.


\section*{CONCLUSIONS}
We have studied how transmission-line (TL) structures, typically used to transmit electromagnetic signals in electronic devices, can effectively transmit vacuum fluctuations between dipoles and drastically enhance their dispersion forces.
This comes about thanks to the unique non-diffractive 1d character of virtual-photon propagation via the TEM mode in TLs. We have shown how the resulting van der Waals (vdW) and Casimir-Polder interactions can become longer-range and larger by orders of magnitude than their free-space counterparts. To this end, we have analytically found, by two independent methods, an expression [equation (\ref{E})] for the dominant TEM-mediated interaction at all inter-dipolar distances, and described how the free-space result is restored at very short distances, by including higher-order modes in the calculations.
Although our approach assumes that the TL is comprised of perfect conductors at zero temperature, it remains accurate for a realistic superconducting coplanar waverguide, for which we have estimated that the enhanced interaction may be directly measured for a \emph{single} pair of superconducting transmons.

We would like to stress the uniqueness of the vdW/Casimir force mediated by a TL as compared to that of other waveguides. The fact that any waveguide allows waves to propagate only in one direction does not guarantee its support of 1d long-range vacuum forces; such modified 1d-like forces are only mediated by the TEM mode that has no cutoff and exists only in TLs. For example, in a hollow metal waveguide, where all transverse modes possess a cutoff, the vdW energy between a pair of dipoles can become extremely short-ranged \cite{MWG}. In a fiber, the fundamental HE$_{11}$ mode might give rise to a more extended vacuum interaction, but since its effective area, unlike that of the 1d-like TEM mode, depends on frequency \cite{OKA}, it is diffractive and does not give rise to the 1d interaction found here.

Our result may pave the way towards the exploration of more complex Casimir phenomena than the simple dipole-dipole non-retarded vdW interaction due to two major effects; i.e., retardation and non-additivity \cite{REV}. As discussed above, both of these may be drastically modified in a 1d geometry, namely, by the presence of a transmission line.
Finally, the predicted modification of the basic interaction between dipoles may prove relevant to diverse areas of applied and basic research: \emph{Circuit QED}, where it can provide a fundamental demonstration of the 1d vacuum effect enabled by such systems; \emph{Quantum Information}, where Rydberg-blockade-based quantum gates may be enhanced; \emph{Classical Electromagnetism}, where the macroscopic dielectric properties of a gas have to be revisited.

\section*{Materials and Methods}

\textbf{QED perturbation theory}
\\
The sum in equation (\ref{E4}) includes 12 different terms, each corresponding to a different set of intermediate states and represented by a diagram \cite{MQED}. E.g. the term corresponding to the diagram in Fig. 2(a) is given by
\begin{eqnarray}
&&-\frac{\hbar^2 c^2}{16 \pi^2 \epsilon^2 A_1A_2}\sum_{n_1,n_2}(\mathbf{d}_{n_1}^{\bot}\cdot\mathbf{d}_{n_2}^{\bot})^2\int_{-\infty}^{\infty} dk \int_{-\infty}^{\infty} dk'
\nonumber \\ &&\times
\frac{|k| |k'|e^{ik z}e^{ik' z}}{(E_{n_1}+\hbar c|k|)(\hbar c|k|+\hbar c|k'|)(E_{n_1}+\hbar c|k'|)}.
\label{E4_3}
\end{eqnarray}
Summing all 12 terms and then performing the integration over $k'$ we arrive at
\begin{eqnarray}
U&=& -\frac{\pi}{2\epsilon^2 A_1A_2}\sum_{n_1,n_2} \frac{(\mathbf{d}_{n_1}^{\bot}\cdot\mathbf{d}_{n_2}^{\bot})^2}{\lambda_{n_1}^2E_{n_1}} F_{n_1n_2}(z),
\nonumber \\
F_{n_1n_2}&=& \int_0^{\infty}dk\sin(2kz)\frac{-2\lambda_{n_1}k^2(k+k_{n_1}+k_{n_2})}{\pi(k_{n_1}+k_{n_2})(k+k_{n_1})(k+k_{n_2})},
\nonumber \\
\label{U12}
\end{eqnarray}
where $k_n=E_n/(\hbar c)=2\pi/\lambda_n$. The integration over $k$ is performed by regularization, yielding
\begin{eqnarray}
&&F_{n_1n_2}(\xi,b)=
\nonumber \\
&&\frac{2b}{b^2-1}\left\{-2\mathrm{Ci}(4\pi\xi)\sin(4\pi\xi)+2b\mathrm{Ci}(4b\pi\xi)\sin(4b\pi\xi)
\right.\nonumber \\
&&\left.-\cos(4\pi \xi)\left[\pi-2\mathrm{Si}(4\pi \xi)\right]+b\cos(4b\pi \xi)\left[\pi-2\mathrm{Si}(4b\pi \xi)\right]\right\},
\nonumber \\
\label{F12}
\end{eqnarray}
where $\mathrm{Ci}(x)$ and $\mathrm{Si}(x)$ are the cosine and sine integral functions, respectively. Here $F_{n_1n_2}(\xi,b)$ is the dimensionless  vdW/Casimir energy contributed by the interaction between the dipolar transition $|g\rangle \rightarrow |n_1\rangle$ of the first dipole and the $|g\rangle \rightarrow |n_2\rangle$ transition of the second dipole, where $\xi=z/\lambda_{n_1}$ and $b=E_{n_2}/E_{n_1}$ represents the asymmetry between the two transitions. Eq. (11) then reveals that the total vdW/Casimir  energy is given by a sum over all such possible interactions between dipolar transitions of the two dipoles. In the vdW and Casimir limits, $\xi\ll 1$ and $\xi\gg 1$, respectively, we obtain,
\begin{eqnarray}
F_{n_1n_2}(\xi\ll1,b)&\approx&\frac{2b\pi}{1+b}
+\frac{16b\pi}{1-b^2}\xi\left[\ln(4\pi\xi)-b^2\ln(4b\pi\xi)\right],
\nonumber \\
F_{n_1n_2}(\xi\gg1,b)&\approx&\frac{1}{8b\pi^3}\frac{1}{\xi^3}.
\label{F12b}
\end{eqnarray}
Expressions (\ref{F12b}) yield the same universal scalings as those of Eq. (\ref{F}) for the two-level dipole case, which is recovered by appropriately taking the limit $b\rightarrow 1$. Alternatively, by directly performing the integration over $k$ in Eq. (11) for a single excited level $|e\rangle$ in each dipole, we obtain equation (\ref{E}) in the main text.
\\\\
\textbf{Scattering of vacuum fluctuations}
\\
The scattered field is proportional to the Green's function of the Helmholtz equation in 1d, equation (\ref{HELM}), which is found to be $\frac{i}{2|k|}e^{i|k||z-z_1|}$. Then, inserting this field into the energy equation (\ref{U}), we arrive at
\begin{equation}
U=\frac{\hbar c}{2\pi\epsilon^2 A_1 A_2}\int_0^{\infty}dk \alpha_1(k)\alpha_2(k) k^2 \sin(2 k z).
\label{UAB}
\end{equation}
Taking the integration on the imaginary axis $k=iu$ in a complex $k$-plane (Wick rotation, assuming the poles of $\alpha_{1,2}(k)$ have some width, i.e. pushing them slightly below the real axis), we obtain
\begin{equation}
U=\frac{-\hbar c}{2\pi\epsilon^2 A_1 A_2}\int_0^{\infty}du \alpha_1(i u)\alpha_2(i u) u^2 e^{-2u z}.
\label{UABW}
\end{equation}
Upon taking the polarizabilities $\alpha_{1,2}$ of a system with a discrete set of transitions as in an atom, $\alpha(k)=\frac{2}{3}\sum_n\frac{E_n|\mathbf{d}_n|^2}{E_n^2-\hbar^2c^2k^2}$ \cite{MQED}, the integration can be performed. The resulting energy $U$, described by special functions, is equivalent to that of Eq. (11) when $|\mathbf{d}^{\bot}_e|^2=(2/3)|\mathbf{d}_e|^2$ is assumed.  Specifically, for the vdW and Casimir limits, we obtain exactly the same analytical expressions of Eqs. (\ref{F12b}) and (\ref{F}).
\\\\
\textbf{Contribution of higher order transverse modes}
\\
In the Supporting Information we calculate, for a coaxial TL, the interaction energy due to the TE and TM modes. We find that the energy contribution of the TE$_{lm}$ and TM$_{lm}$ modes with cutoff frequency $c k_{lm}$ scales like $K_0(k_{lm}z)$ and $e^{-k_{lm}z}$ respectively, where $K_0(x)$ is the modified Bessel function. Since $k_{lm}>\pi/a$, at long distances $z>a$ both decay exponentially and are negligible with respect to the TEM mode energy. However, at short distances $z\ll a$, we numerically verify that the dominant TM modes sum up to give exactly the free-space interaction. This was also recently shown for the dispersion interaction in a metallic waveguide \cite{MWG}, and we expect it to hold for all TLs. Namely, for distances $z\gg a$ we can indeed consider only the TEM mode, whereas the free-space result is recovered for $z\ll a$, owing to the role of other transverse modes.
\\\\
\textbf{Scaling of $A(x,y)$ with $a$}
\\
Considering a coaxial line for example [Fig.1(a)], upon normalizing the electric field of the TEM mode, we find $\sqrt{A}=\sqrt{2\pi \ln(b/a)}\rho$, with $\rho=\sqrt{x^2+y^2}$ \cite{POZ}. Since $a<\rho<b$, taking e.g. $b=2a$ and $\rho=a$ we obtain $\sqrt{A}\approx 2.1a$, such that $A\sim a^2$.

\begin{acknowledgments}
The support of ISF, BSF, the Wolfgang Pauli Institute and the FWF (Project No. P25329-N27) is acknowledged. We appreciate fruitful discussions with Yoseph Imry and  Grzegorz {\L}ach.
\end{acknowledgments}

\end{document}